\documentclass{ws-rv975x65}
\usepackage{epsfig}
\usepackage{subfigure}     
\usepackage{ws-rv-van}     
\makeindex
\begin{document}

\chapter[Using World Scientific's Review Volume Document
Style]{Frustration and Fluctuations in Systems with Quenched Disorder}
\label{ra_ch1}

\author[D.L.~Stein]{D.L.~Stein\footnote{Email address: {\tt daniel.stein@nyu.edu}}}

\address{Department of Physics and Courant Institute of Mathematical Sciences,\\
New York University,\\
4 Washington Pl.,\\
New York, NY 10003 USA}

\begin{abstract}
  As Phil Anderson noted long ago, frustration can be generally defined by
  measuring the fluctuations in the coupling energy across a plane
  boundary between two large blocks of material. Since that time, a number of
  groups have studied the free energy fluctuations between (putative)
  distinct spin glass thermodynamic states. While upper bounds on such
  fluctuations have been obtained, useful lower bounds have been more
  difficult to derive.  I present a history of these efforts, and
  briefly discuss recent work showing that free energy fluctuations between
  certain classes of distinct thermodynamic states (if they exist)
  scale as the square root of the volume. The perspective offered
  here is that the power and generality of the Anderson conception of
  frustration suggests a potential approach toward resolving some 
  longstanding and central issues in spin glass physics.
\end{abstract}

\body

\section{Phil Anderson and Spin Glass Theory}
\label{sec:intro}

It is a great pleasure, both personally and scientifically, to contribute
to this volume in honor of Phil Anderson's 90th birthday. The importance
and influence of Phil's research in shaping the modern field of condensed
matter physics (including coining the term, along with Volker Heine) is
widely recognized. There are few currently active, fruitful areas of
condensed matter research that have not been either created or (at least)
strongly influenced by Phil. His influence, moreover, is not limited to
condensed matter physics: he pointed the way~\cite{Anderson62} toward what
is now universally known as the Higgs mechanism; set the stage for later
developments in complexity science by emphasizing the importance of a
nonreductionist scientific viewpoint~\cite{Anderson72} (not a widely held
view at the time); and later explored and emphasized the connections
between the statistical mechanics of quenched disorder and problems in
biology, graph theory, and other areas outside of
physics~\cite{Anderson88}.

It is the topic of quenched disorder that I will address here. I will not
discuss its applications to other areas, but will instead return to the
theory's roots. The subject of disordered systems --- in particular glasses
and spin glasses --- has been a longstanding interest of Phil's, and one in
which he has made numerous fundamental contributions.  At the height of his
interest, he identified the problem of understanding the physics of
structural glasses and their magnetic counterparts, spin glasses, as one of
the central unsolved problems in condensed matter
physics~\cite{Anderson79}. His views may well have changed in the ensuing
decades, but one can safely argue that understanding the effect of quenched
randomness on the condensed state has presented one of the most persistent
and confounding set of problems in modern condensed matter physics.

On the subject of spin glasses proper, Phil's contributions are too
numerous to list, but it would be a dereliction of duty not to mention two
of his most foundational papers. The first is well-known: his 1975
paper~\cite{EA75} with Sam Edwards that swept away numerous distracting
details and identified quenched, conflicting ferromagnetic and
antiferromagnetic interactions as the ultimate microscopic basis of spin
glass behavior. Edwards and Anderson (hereafter referred to as EA) used
this idea to propose a simplified model Hamiltonian that has
since formed the basis of most theoretical investigations (we include here
studies of the Sherrington-Kirkpatrick model~\cite{SK75}, which is an
infinite-range version of the EA model).  The other important idea proposed
in the EA~paper concerned the nature of the spin glass order parameter, but
that's less relevant to the discussion below.

The second of these papers, not as widely known or cited, concerns Phil's
joint (with G{\'e}rard Toulouse) introduction of the concept of
frustration.  The story goes that G{\'e}rard attended a lecture in 1976 in
which Phil wrote on the blackboard, ``The name of the game is
frustration.''  Whether this was elaborated on in the talk I couldn't say,
and the principals will have to provide the details --- if they
remember. But it would have been characteristic of Phil to make this
cryptic remark without elaboration and move on. Inspired, G{\'e}rard
published a classic paper the next year~\cite{Toulouse77} that remains the
canonical definition --- both conceptual and operational --- of
frustration. In this formulation, one considers the spins at lattice sites
that form a closed loop on a lattice.  If there are an odd number of
antiferromagnetic couplings on the edges constituting the loop, the spins cannot be
arranged to satisfy all of the interactions. 

The following year, Phil published an alternative, and more general,
definition of frustration~\cite{Anderson78} in which one studies the free
energy fluctuations of two blocks of material (glass, spin glass,
ferromagnet, what have you) that have independently relaxed to their
respective ground states. I will elaborate both on this and the Toulouse
definition of frustration in Sect.~\ref{sec:frustration}.  For now I will
just note that this latter approach has not received as much attention as 
Toulouse's, but nonetheless --- as usual for Phil --- it is enormously
prescient. In fact, I will argue below that this alternative approach to
frustration may, after a long period of dormancy, contain just the right
perspective to resolve some longstanding open questions at the heart of
spin glass physics.

But for now, the main point is that in both cases frustration arises when a
system contains many fixed, conflicting internal constraints, not all of
which can be simultaneously satisfied. Like art, physics can sometimes
imitate life.

The final topic of this informal account concerns Phil's role in the naming
of spin glasses. Phil didn't name them himself, although he clearly was
involved~\cite{Anderson77}. Regardless of the details, he was as usual
present at the creation, and provides an amusing discursion on the early
days.  Most accounts, including Phil's, credit Bryan Coles with inventing the
term ``spin glasses'', although details among the accounts vary somewhat. 
A competing (and in my opinion, more ungainly) term
that had gained some traction at the time was ``mictomagnetism'', and we
might well today be referring to mictomagnets rather than spin
glasses.\footnote{In fact, the term still survives, and you can google it,
  although Google will insist that you must have meant ``micromagnets.''  Don't
  give in.} Phil's dry sense of humor led to a disquisition on the
etymology of the term, and one cannot do justice to his description other
than quoting it in full~\cite{Anderson77}:

``A few weeks ago I received a letter from Ralph Hudson of the NBS
objecting to this term, on the basis that he thought that the only other
word in the English language using the same root was `micturation' and the
root was Latin for `urine'. I think myself that the term is very
descriptive: back in the Middle West we used to refer to something as
`p---poor' if it was not worth anything more substantial, and that is a
good description of this kind of magnetism. The remanence is small and
sluggish, there are peculiar training phenomena, and the susceptibility is
often very history dependent. Unfortunately, I am assured by Collin Hurd
and by the OED that Hudson is incorrect and that `micto' is a legitimate
Greek root meaning `mixed'. ''

Those interested in Phil's first-hand perspective of the early days of spin
glass research should consult his series of Physics Today ``reference frame'' 
articles~\cite{AndersonPT1,AndersonPT2,AndersonPT3,AndersonPT4,AndersonPT5,AndersonPT6,AndersonPT7}
that helped bring the subject to the attention of the broader physics
community.

\section{Free Energy Fluctuations in the Random Field Ising Model}
\label{sec:fluctuations}

Systems with quenched disorder possess several features distinguishing them
from homogeneous systems. Our focus is on one of these: their energies and
free energies are random variables depending on the disorder, and their
concomitant fluctuations contain information that can potentially resolve
central open questions that remain intractable to this day. These questions
include the conjectured multiplicity of pure and ground thermodynamic
states, the relations between such distinct states (if they exist), the
geometry and energy scaling of their relative interfaces, and so on.

To illustrate the potential usefulness of the information provided by these 
fluctuations, we turn briefly to a different system: the
random-field Ising model~(RFIM), which is a uniform Ising ferromagnet subject to a
random external field. It can be modelled using the Hamiltonian
\begin{equation}
\label{eq:rfim}
{\cal H}_h=-\sum_{\langle x,y\rangle}\sigma_x\sigma_y-\epsilon\sum_xh_x\sigma_x\, .
\end{equation}
Here $x$ and $y$ are lattice sites in the $d$-dimensional cubic lattice
${\bf Z}^d$, $\sigma_x=\pm 1$ is an Ising spin at site $x$, the first sum
is over nearest-neighbor pairs of sites only, and the fields $h_x$ are
independent, identically distributed random variables representing local
external fields acting independently at each site $x$. For simplicity, we
take the probability distribution of the $h_x$'s to be Gaussian with mean
zero and variance~one. The subscript $h$ on the LHS of~(\ref{eq:rfim})
refers to a particular realization of the $h_x$'s.

One can now ask whether the ferromagnetic ground state is unstable to
breakup by the random field.  The answer is clearly yes if $\epsilon$ is
sufficiently large. But is it true for {\it any\/} nonzero $\epsilon$?

For a uniform field this question is trivial: in any dimension, a field of any fixed, 
nonzero magnitude determines the magnetization direction at all temperatures, 
and so there is no phase transition. A
simple scaling argument explains why. In zero field below $T_c$, consider
the positively magnetized (i.e., ``up'') phase. Now apply a small uniform
field of magnitude $h$ pointing down.  Overturning a compact patch (or ``droplet'')  of spins
of length scale $L$ to align with the field is energetically favorable for
sufficiently large $L$: for Ising spins, the cost in surface energy is of
order $L^{d-1}$ while the lowering of bulk energy is of order $hL^d$. So,
no matter how small $h$ is, the system can lower its energy by overturning
a sufficiently large droplet. At positive temperature,
overturning droplets of spins is certainly entropically favorable as
well. Consequently, in any dimension and at any temperature (including zero), there is a unique
Gibbs state in an external uniform field of any nonzero magnitude.

Now consider the case when quenched disorder is present, i.e., when the field is
random.  This requires a modification of the above argument, which was
provided in 1975 by Imry and Ma~\cite{IM75}. The boundary energy, which
depends only on the ferromagnetic couplings, is unchanged. The bulk energy,
however, is determined by the fields, which now fluctuate from region to
region. Nevertheless, in an arbitrary large droplet containing $L^d$ spins,
the central limit theorem requires that the typical bulk energy scales as
$L^{d/2}$. So in two dimensions the competing boundary and bulk energies scale similarly with
volume.  Imry and Ma concluded that in two dimensions and below, the
ferromagnetic ground state should be unstable for any nonzero
$\epsilon$, while above two dimensions (and with small but nonzero $\epsilon$) 
ferromagnetic long-range order persists.  (In fact, Imry and Ma mainly focused on
continuous spin models, where the boundary energy scales as
$L^{d-2}$, giving a lower critical dimension of 4.)

Normally, such an argument would be sufficient to settle the matter, but a
few years later a more detailed field-theoretical analysis based on
supersymmetry~\cite{PS79} concluded that the critical behavior of a
$d$-dimensional spin system in a random external field is equivalent to
that of the corresponding $(d-2)$-dimensional system in the {\it absence\/}
of an external field. This ``dimensional reduction'' argument therefore 
predicts the lower critical dimension of the RFIM to be three.

This controversy was eventually resolved by rigorous mathematical
arguments, first by Imbrie~\cite{Imbrie85} and later by Aizenman and
Wehr~\cite{AW90,WA90}. Imbrie proved that the Ising model in a random
magnetic field in three dimensions exhibits long-range order at zero
temperature and sufficiently small disorder, indicating that the lower
critical dimension of the RFIM is indeed two. Aizenman and Wehr later
proved that in two dimensions at all temperatures and fields, the RFIM
possesses a unique Gibbs state.

This now ancient controversy is recounted for two
reasons. First, it represents an interesting --- and rare --- example where
rigorous mathematics resolved an open and important controversy in
theoretical (and indeed, experimental) physics. It may be the case that
something similar will be required for spin glasses and possibly even
structural glasses, where the most basic questions have persisted as a
subject of intense controversy over decades.

More relevant to this paper, though, is that Aizenman and Wehr essentially
made the Imry-Ma argument rigorous by analyzing fluctuations (with respect
to the quenched disorder) of the free energy difference between putative
positively and negatively magnetized states. Although the RFIM is not 
a frustrated system, the spirit of the Aizenman-Wehr method
aligns with the Anderson approach to characterizing and understanding
frustration.

\section{Frustration}
\label{sec:frustration}

We turn now to finite-dimensional spin glasses, where almost all of the
basic questions remain open.  These include whether an equilibrium phase
transition occurs above some dimension; if so, the nature of the broken
symmetry (if any) of the spin glass phase; whether (up to global symmetry
transformations) the spin glass phase is unique; if not, the nature of the
relationships among the many spin glass phases; whether there exists an
upper critical dimension above which mean field theory holds;\footnote{In
  homogeneous systems, this question usually refers only to behavior at or
  near the critical point.  For spin glasses it is considerably
  more far-reaching. Here we're asking whether the
  {\it low\/}-temperature properties ---- i.e., the order parameter and the
  nature of broken symmetry --- corresponds in {\it any\/} finite dimension to the
  replica symmetry
  breaking~\cite{Parisi79,Parisi83,MPSTV84a,MPSTV84b,MPV87} that occurs in
  the low-temperature phase of the infinite-range Sherrington-Kirkpatrick
  model~\cite{SK75}. This is generally not an issue in homogeneous systems, where
  mean-field theory usually provides a useful guide to the nature of the
  low-temperature phase well below $T_c$ in any dimension where a phase
  transition occurs.} and numerous others. And this list doesn't include
questions concerning the nonequilibrium dynamical behavior of
spin glasses, which won't be addressed in this paper.

For concreteness we confine our attention to nearest-neighbor models
defined by the EA~Hamiltonian~\cite{EA75}:
\begin{equation}
   \label{eq:EA}
   {\cal H}_{\cal J}=-\sum_{\langle x,y\rangle} J_{xy} \sigma_x\sigma_y -h\sum_x\sigma_x\ ,
   \end{equation}
   where $x$ and $y$ are sites in the $d$-dimensional cubic lattice,
   $\sigma_x=\pm 1$ is the Ising spin at site $x$, the couplings $J_{xy}$
   are independent, identically distributed random variables, ${\cal J}$
   denotes a particular realization of the couplings (corresponding
   physically to a specific spin glass sample with quenched disorder), $h$
   is an external magnetic field, and the first sum is over nearest
   neighbor sites only.  We hereafter take $h=0$ and the spin couplings
   $J_{xy}$ to be symmetrically distributed about zero; consequently, the
   EA Hamiltonian in~(\ref{eq:EA}) possesses global spin inversion
   symmetry.
   
   A striking feature of the EA~Hamiltonian is the
   presence of {\it frustration\/}, meaning the inability of any spin
   configuration to simultaneously satisfy all couplings.  It is easily
   verified that, in any dimension larger than one, all of the spins along
   any closed circuit ${\cal C}$ in the edge lattice cannot be
   simultaneously satisfied if
   \begin{equation}
   \label{eq:frustration1}
   \prod_{\langle x,y\rangle\in{\cal C}}J_{xy}<0\, .
   \end{equation}
   This definition of frustration is due to Toulouse~\cite{Toulouse77}.

   Toulouse's geometry-based definition is appealing on several levels, and
   has been the starting point for numerous investigations (see, for
   example, Refs.~[27,28]\nocite{FHS78,BF86}).  It provides a simple test to
   determine whether a given type of spin system possesses frustration, and
   suggests the underlying reason why certain systems may possess multiple
   pure or ground states.  More generally, the quantification of
   conflicting internal constraints provides a powerful conceptual tool for
   understanding certain general aspects of complex behavior in the broader
   study of complex systems.

Its drawback is that in some sense the Toulouse approach is {\it too\/}
well-suited to spin glasses; it is difficult to see how it can be
generalized in a natural way to non-spin systems that surely possess
frustration, such as structural glasses or combinatorial optimization
problems. For these systems the Anderson definition of frustration is more
useful; it is sufficiently general that (with minor modification as
needed) it should apply to any system. As we will
see, it also provides a conceptual starting point for mathematical studies
that hold promise for resolving the long-controversial issue of pure/ground
state multiplicity in frustrated systems with quenched disorder.

The idea itself is rather simple, although its simplicity conceals a
profound and very useful insight. Based on a preliminary study of Anderson
and Pond~\cite{AP78}, Anderson proposed~\cite{Anderson78} considering the
free energy fluctuations of two statistically identical blocks of the same
material. It is simplest to describe the procedure at zero temperature,
although it is easily modified for positive temperature. So let each block
of material independently relax to its ground state\footnote{For a finite
  system with specified boundary conditions, such as periodic, and
  continuous disorder, such as Gaussian, the ground state is unique up to a
  global symmetry.}.  One then brings the two blocks together and measures
the fluctuations in the coupling energy across their interface. In
nonfrustrated systems, such as ferromagnets, the energy fluctuations 
scale as the surface area $A$ of contact. This scaling holds for both
homogeneous and random ferromagnets (in which the bond strengths are
positive, i.i.d.~random variables).

However, if frustration is present, then it will be the case that
\begin{equation}
\label{eq:frustration2}
\lim_{A\to\infty} \langle E^2\rangle/A^2=0\, .
\end{equation}
In fact, one can turn this around and use~(\ref{eq:frustration2}) as the
general definition of frustration, which we will hereafter do.

Using reasoning based on the central limit theorem Anderson further conjectured that
\begin{equation}
\label{eq:frustration3}
\lim_{A\to\infty} \langle E^2\rangle/A=O(1)\, .
\end{equation}
That is, in a frustrated system one might expect the energy fluctuations of
the ground states to scale as the {\it square root\/} of the surface
area of contact.  However, (\ref{eq:frustration3}) is a rough estimate.

The definition~(\ref{eq:frustration2}) of frustration, although not as
widely known or appreciated as~(\ref{eq:frustration1}), contains the
seeds of a powerful approach to understanding realistic spin glasses and
other complex systems. We turn now to a natural outgrowth of this approach,
namely, scaling theories of the spin glass phase.

\section{Scaling Theories of the Spin Glass Phase}
\label{sec:scaling}

The idea of investigating fluctuations of
free energy differences in the presence of frustration 
leads naturally to a scaling approach for
understanding the low-temperature spin glass phase. This approach, which
has a long history in the study of phase transitions and broken symmetry,
examines how the ``stiffness'' of the low-temperature phase scales with the
system size $L$. A stable phase requires the stiffness --- roughly
speaking, the free energy cost associated with overturning a droplet 
of spins --- to increase (or at least not decrease) with $L$, usually as a power law, although other forms are possible in
principle.

This approach as applied to spin glasses began with the early work of
Anderson and Pond~\cite{AP78}, and was developed throughout the
1980's~\cite{BC82,BC83,Mac84,CB85,BM85,FH86,BM87,FH88a,FH88b}. The
essential idea is to study the fluctuations in a finite volume of the spin
glass free energy as one changes boundary conditions, for example from
periodic to antiperiodic. Physically, such a change in boundary conditions
generates relative interfaces inside the box, so one is effectively
studying the interface free energy. Given the close relation between this
approach and that of the Anderson definition of frustration, one might
expect that the presence of frustration will generate profound effects on
such interfacial free energies. And of course it does.

\subsection{Interface Geometry}
\label{subsec:geometry}

Before proceeding, some remarks are necessary concerning the relation
between this procedure and the presence of many states. While switching
from periodic to antiperiodic boundary conditions always generates relative
interfaces, geometrically one of three things can happen.  Consider a
``window''~\cite{NS98,NS03b} of large but fixed linear size~$w$ centered at
the origin, and consider the interfaces generated when $L\gg w$. The first
possibility is that as $L$ grows increasingly larger (with $w$ fixed), the
interfaces eventually move outside of the window, so that the thermodynamic
state {\it inside\/} the window is the same for both periodic and antiperiodic boundary
conditions. If this is the case, then there are only two spin glass pure
states (or ground states at zero temperature), which are global flips of
each other.

The other possibility, of course, is that no matter how far away the
boundaries move, interfaces always penetrate inside the window. This is the
signature of multiple spin glass pure state pairs. This possibility can be
further divided into two parts: either the interfaces have vanishing
density as the window size increases (with the order of limits being
$L\to\infty$ followed by $w\to\infty$) or else the interface density
remains bounded away from zero. The former zero-density case is what one
finds in the ferromagnet, which exhibits $d-1$-dimensional~interfaces in a
$d$-dimensional system. In contrast, the latter ``space-filling'' case, if
it occurs, requires $d$-dimensional interfaces within a $d$-dimensional
system, which would signify a novel feature of spin glasses. Huse and
Fisher~\cite{HF87,FH87} refer to the zero-density situation as ``regional
congruence'': the states are locally the same almost everywhere. The more
interesting situation with space-filling interfaces was denoted
``incongruence'': the states, although similar in a statistical sense, are
dissimilar everywhere. It was proven in~\cite{NSregcong01} that any
procedure using boundary conditions chosen in a coupling-independent manner
(as in the periodic-antiperiodic situation above) always results either in
a single pair of states (no interfaces in the window)\footnote{An absence
  of interfaces within the window implies that all interfaces necessarily
  generated by changing boundary conditions must be zero-density, since
  positive density interfaces {\it must\/} penetrate the
  window\cite{NS02}. The difference between this case and the regionally
  congruent case is that in the latter, zero-density interfaces continue to
  penetrate the window no matter how far away the boundaries are, while in
  the former, the interfaces deflect to infinity as the boundaries move out
  to infinity.} or else many incongruent pairs of states.  Regionally
congruent states, should they exist, can only be generated using
coupling-{\it dependent\/} boundary conditions, requiring procedures as yet
unknown. In what follows we therefore confine the discussion to incongruent
states.

\subsection{Interface Energetics}
\label{subsec:energetics}

The preceding discussion focuses exclusively on the geometry of interfaces
between pure states; we now discuss their energetics. In a ferromagnet,
whether homogeneous or random, all couplings have the same sign, so the
interface energy scales with the number of edges it comprises. In a spin
glass, interface energetics remain an open problem, and a very important
one: if one knows how the interface energy scales with its size, one can
finally resolve the longstanding open question regarding the multiplicity
of pure states in the spin glass phase.

Here's why.  By the same reasoning that led to the Anderson definition of
frustration, it is reasonable to expect that the free energy of a
space-filling interface of linear extent $L$ should scale no faster than
$L^{d/2}$. However, it is possible that correlations could lower the
minimal interface free energy, so that it scales as $L^\theta$, with $0\le\theta\le d/2$
(assuming a stable spin glass phase). The lower bound of $\theta=0$ is
predicted\footnote{More precisely, this scaling applies to interfaces between pairs
of incongruent states within the {\it same\/} thermodynamic state. Interfaces between incongruent states
belonging to {\it different\/} thermodynamic states, which would be expected within the RSB 
picture upon switching from periodic to antiperiodic boundary conditions, would presumably have
$\theta>0$. For a discussion of thermodynamic states within RSB, see~[46], Sect.~7.9.} by the mean-field replica symmetry breaking~(RSB) picture of the
spin glass phase~\cite{Parisi79,Parisi83,MPSTV84a,MPSTV84b,MPV87}, while
$0<\theta<(d-1)/2$ is predicted by the chaotic pairs
picture~\cite{NS96c,NS03b,NSbook}. Both are many-states pictures, though
with very different thermodynamics and organization of the incongruent pure
states, as will be discussed below.

Based on numerical results and scaling arguments, Fisher and
Huse~\cite{FH86} conjectured a stronger upper bound than $L^{d/2}$; they
argued that in fact $\theta\le (d-1)/2$. If this is correct, then the
question of whether incongruent states exist reduces to the question of
whether their interface free energy should scale faster or slower than
$L^{(d-1)/2}$. Fisher and Huse argued, along roughly similar lines to
Anderson, that it should scale as $L^{d/2}$. The resulting contradiction between the
upper and lower bounds leads to the conclusion that incongruent states
cannot exist in the EA~spin glass in any finite dimension.

But is the conjectured upper bound $\theta=(d-1)/2$ correct? This was
initially a matter of some controversy, but within a few years it was
proved, using rigorous mathematical arguments, by Aizenman and
Fisher~\cite{AFunpub}, and independently and a little later by Newman and
Stein~\cite{NSunpub}.  Unfortunately, neither was ever published, but the
arguments are now familiar to those who work in this area. I will
informally sketch the basic idea of the proof here.  First, let me state
the exact result more formally:

\begin{theorem}
  Let $F_P$ be the free energy of the finite-volume Gibbs state generated
  by Hamiltonian~(\ref{eq:EA}) (with $h=0$) in a box $\Lambda$ of volume
  $L^d$ using periodic boundary conditions, and $F_{AP}$ that generated
  using antiperiodic boundary conditions. Let $X_\Lambda = F_P -
  F_{AP}$. Then ${\rm Var}(X_\Lambda)\le{\rm const.}\times L^{d-1}$, where
  ${\rm Var}(\cdot)$ denotes the variance over all of the couplings inside
  the box.
\end{theorem}

Physically, this implies that the fluctuations in free energy (and
therefore the interface free energy) induced by changing boundary
conditions from periodic to antiperiodic in a box of volume $L^d$ scales as
$L^{(d-1)/2}$. The theorem as stated above is more restrictive than necessary;
the same result applies for any two boundary conditions that are {\it
  gauge-related\/}, i.e., that can be transformed into each other by
reversing the sign of some subset of couplings on the boundary of the
box.\footnote{A coupling on the boundary of the box connects a site inside
  $\Lambda$ to one on the boundary of $\Lambda$.} So, for example, the same
result holds for any two distinct fixed boundary conditions.

  The proof uses a {\it martingale decomposition\/} of the free energy
  difference, as follows.  First note that, by the gauge-relatedness of the two boundary conditions,
  $E[X_\Lambda]=0$, where $E[\cdot]$ denotes a full average over all the couplings inside the box. Next 
  number each coupling inside the box, as shown in
  Fig.~1.
\begin{figure}[th]
\label{fig:martingale}
\centerline{\epsfig{file=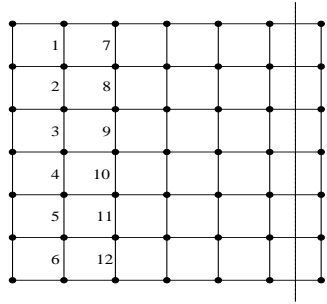,width=4.2cm,height=4.0cm}}
\caption{Schematic of a box with each coupling numbered (only the first
  twelve are explicitly indicated). The dashed line on the right indicates
  couplings whose transformation $J_{xy}\to -J_{xy}$ is equivalent to
  switching from periodic to antiperiodic boundary conditions, and which
  according to the argument in the text are the only ones contributing to 
  the variance.}
\end{figure}
We now successively average the free energy difference $X_\Lambda$ over an
increasing number of couplings. Let $x_{\Lambda,j}=E[X_\Lambda|b_1b_2\ldots
b_j]$, where $E[A|y_1,y_2,\ldots y_k]$ is the conditional expectation, or
average, of $A$ conditioned on the random variables $y_1,y_2,\ldots
y_k$. That is, suppose that $A$ is a random quantity depending on the $N$
random variables $y_1,y_2,\ldots y_N$. Then $E[A|y_1,y_2,\ldots y_k]$
represents the quantity resulting from averaging $A$ over
$y_{j+1},y_{j+2}\ldots y_N$. So $x_{\Lambda,0}$ represents $X_\Lambda$
fully averaged over all of the couplings in the interior of the box, and
$x_{\Lambda,N}=X_\Lambda$, the original unaveraged free energy difference.

If there are $N$ couplings inside the box, it is not hard to see that
\begin{equation}
\label{eq:martingale}
{\rm Var}(X_\Lambda)={\rm Var}\Bigl[\sum_{j=0}^{N-1}\Big(x_{\Lambda,j+1}-x_{\Lambda,j}\Big)\Bigr]=\sum_{j=0}^{N-1}{\rm Var}\Bigl(x_{\Lambda,j+1}-x_{\Lambda,j}\Bigr)\, ,
\end{equation}
where the second equality follows because the so-called martingale
differences $x_{\Lambda,j+1}-x_{\Lambda,j}$ are orthogonal quantities.\footnote{Two
  random variables $A$ and $B$ are orthogonal if $E[AB]=0$.  It is easy to
  see that this holds for any two martingale differences.}

Now suppose that the antiperiodic boundary conditions are applied to the
right and left boundaries, and consider an $x_{\Lambda,j}$ conditioned on
any subset of the couplings {\it except\/} those cut by the dashed line in
Fig.~1.  Any such $x_{\Lambda,j}=0$, because one is averaging over all of
the couplings cut by the dashed line, and taking $J_{xy}\to -J_{xy}$ over
these couplings is equivalent to switching between periodic and
antiperiodic boundary conditions. The only nonzero contributions to the sum
in~(\ref{eq:martingale}) therefore comes from conditioning on these
boundary couplings, each of which contributes a term of order one to the
variance. This completes the argument.

\section{Interface Free Energy Fluctuations}
\label{sec: lb}

If we can now find a strong {\it lower\/} bound for interface free
energies between incongruent states, we would be in a position to determine
whether such states can exist at all; and if they do, what their properties
should be. Unfortunately, finding a lower bound is considerably more
difficult than finding an upper bound, for reasons that will be discussed
below. It has been almost 25 years since the upper bound was proved, and no
progress on finding a lower bound has been made until very recently.

Unlike the upper bound, which is scenario-independent, construction of a
lower bound requires a specific picture of the spin glass phase. In other
words, one first needs to ask: what's doing the fluctuating? We will
examine here four distinct pictures that have been proposed, each of which gives a different answer 
to the question. 

\subsection{Scenarios for the Low-Temperature Spin Glass Phase}
\label{subsec:scenarios}

Probably the most familiar are the {\it mixed-state\/} pictures, in which
the spin glass phase consists of infinitely many thermodynamic states, each
of which is itself a mixture of infinitely many incongruent pure state pairs with
nonzero weights (within their respective thermodynamic states). The
mean-field-inspired RSB scenario is such a picture, although others are
also possible. Recall from Sect.~\ref{subsec:energetics} that in these
scenarios the smallest interface free energies remain order one independently of the
interface size ($\theta=0$).

Almost equally familiar are the scaling/droplet pictures discussed in
Sect.~\ref{sec:scaling}. These are two-state pictures with $\theta>0$ and
in which no interface appears in a window far from the boundaries when one
switches from periodic to antiperiodic boundary conditions.

There are two other less familiar pictures that should nevertheless be
included: the ``TNT'' picture of Krzakala-Martin~\cite{KM00} and
Palassini-Young~\cite{PY00}, and the chaotic pairs picture discussed in
Sect.~\ref{subsec:energetics}. If one constructs a $2\times 2$~grid listing
the different possibilities for interface geometry (space-filling
vs.~zero-density) and energetics ($\theta=0$ vs.~$\theta>0$) then these
additional pictures are required for
completeness~\cite{NS03a}.\footnote{They also arise naturally from a
  metastate analysis of possible spin glass phases~\cite{NS96c,NS97,NSBerlin,NS03b}.} The TNT~picture is presumably a
two-state picture~\cite{NSregcong01} with zero-density interfaces and
$\theta=0$, while chaotic pairs is a many-state picture with space-filling
interfaces and $\theta>0$. Unlike RSB, it is not a (nontrivial) mixed-state
picture: while it contains infinitely many distinct thermodynamic states,
each one consists of a single spin-reversed pure state pair.

So of these four, only RSB and chaotic pairs possess incongruent states,
and any lower bound on interface free energies can help to determine
whether they are likely candidates for the spin glass phase, or else are
not allowed at all. How so?  Consider again the energetics of interfaces
between putative incongruent pure states. As already mentioned, unlike in
the ferromagnet, whether homogeneous, random-bond, or random-field, the
sign of the free energy difference between two putative spin glass states
varies as one moves along their relative interface.  If the single-coupling
energy differences are independent, then one expects an energy that varies
as the square root of the number of couplings in the interface --- i.e., in
a volume of size $L$, the fluctuations in the interface free energy would
scale as $L^{d/2}$, as in the conjecture accompanying the Anderson
definition of frustration. If this were indeed the case, such fluctuations
would violate, in any finite dimension, the upper bound of $L^{(d-1)/2}$
described in Sect.~\ref{subsec:energetics}, and the existence of
incongruent states would therefore be impossible in finite-dimensional spin glasses.

But the single-coupling free energy differences are certainly {\it not\/}
independent. Whether the correlations in free energy fluctuations as one
moves along the interface are strong enough to decrease the free energy scaling exponent
from its independent value is the crucial question that determines whether
incongruent states are present in spin glasses --- or not. Hence,
determining an accurate lower bound is of central importance for resolving
the question of multiplicity of states in realistic spin glasses.

\subsection{Lower Bound}
\label{subsec:lb}

If it's so important, why has it taken so long to find a lower bound for
free energy difference fluctuations? Progress has been held back by several
technical hurdles; the two most troublesome are the ``cancellation
problem'' and the ``identification problem''.

The cancellation problem has already been discussed in
Sect.~\ref{subsec:energetics}. If incongruent states exist, then an
arbitrarily chosen coupling will have, with positive probability, a free
energy difference of order one between the two states. But if one uses the
usual techniques, such as martingale differences, to extrapolate to the
entire volume, cancellations between the many terms lead to an ambiguous
outcome.

Equally difficult is the identification problem. It is hard to see how one
can estimate the extent of free energy difference fluctuations between two
thermodynamic states $\Gamma$ and $\Gamma'$ without averaging over the
couplings inside the volume. But as one does so, what happens to the
original states? Unlike in ferromagnets and other homogeneous systems, there is no
clearcut connection between boundary conditions and thermodynamic states,
and (if there are many incongruent states) the states themselves can change as one varies the couplings
inside the box. So {\it a priori\/} it is not even clear what one is
calculating during the averaging procedure.

In a very recent paper~\cite{ANSW14}, these and other problems were finally
surmounted, although (at the moment) for a limited class of incongruent
states. For these incongruent states it was found that the fluctuations in free energy
differences do indeed scale as $L^{d/2}$; or more formally, the variance of
the free energy difference between the incongruent states considered
scales linearly with the volume.

If these results can be extended to the set of incongruent states in
general, does one then have a contradiction? In a strict mathematical
sense, not quite yet.\footnote{Whether such a result already provides a sufficient
  heuristic or theoretical physics-style argument for the nonexistence of
  incongruent states is left up to the reader.} The problem is that the
quantity for which the lower bound is derived is not exactly the same as
that for which the upper bound was derived, although both are just different
representations of the free energy difference in a finite volume, and so are
equivalent in a physical sense.

The upshot is that we may be on the cusp of resolving the problem of
multiplicity of pure states in realistic spin glasses --- but at the moment
it is unclear as to whether the results can be extended to bring the upper
and lower bounds into alignment. Whether this is eventually done or not, it
is clear that the insights that Phil Anderson had thirty-five
years ago into the nature of frustration are still actively guiding and
influencing fundamental research today.

\section*{Acknowledgements}
The author thanks Louis-Pierre Arguin and Chuck Newman for useful comments
on the manuscript, and Louis-Pierre Arguin, Chuck Newman, and Janek Wehr
for an interesting and enjoyable collaboration that led to the work
described in Section~{\ref{subsec:lb}}.  This research has been supported
in part by U.S.~National Science Foundation Grants~DMS~1207678 and
OISE~0730136.

\bibliographystyle{ws-rv-van}
\bibliography{refs}
                    
\end{document}